\documentclass[12pt]{article}
\usepackage{graphicx}
\usepackage{geometry}
\usepackage{mdwlist}
\usepackage[colorlinks=true, urlcolor=blue]{hyperref}

\makeatletter
\g@addto@macro\@verbatim\footnotesize
\makeatother

\title{ ScalienDB: Designing and Implementing \\
        a Distributed Database using Paxos }
\author{ Marton Trencseni, \texttt{mtrencseni@gmail.com} \and
		 Attila Gazso, \texttt{agazso@gmail.com}
}
\date{}

\begin{document}

\maketitle

\abstract{ ScalienDB is a scalable, replicated database built on top of the Paxos algorithm. It was developed from 2010 to 2012, when the startup backing it failed. This paper discusses the design decisions of the distributed database, describes interesting parts of the C++ codebase and enumerates lessons learned putting ScalienDB into production at a handful of clients. The source code is available on \href{https://github.com/scalien/scaliendb}{Github} under the AGPL license, but it is no longer developed or maintained. }

\section{ Introduction }
%%%%%%%%%%%%%%%%%%%%%%%%

Scalien was a NoSQL startup based in Budapest, Hungary, founded by the two authors in 2009. The company operated until the end of 2012, when the company failed because we were unable to secure venture capital.  Scalien's first database technology was called Keyspace and was written in 2009 \cite{Keyspace}. In 2010, building on the Keyspace experiences, we started working on ScalienDB after receiving minimal funding in the form of a development contract from a client.

Our goal with ScalienDB was to build a NoSQL product that has strong consistency guarantees about data similar to traditional SQL databases, or at least stronger guarantees than other NoSQL databases. This was to be ScalienDB's unique selling point. In retrospect Scalien suffered from a severe lack of product-market fit. NoSQL databases were called upon in use-cases when consistency guarantees were not that important. We also had a hard time communicating the strong points of ScalienDB, as these are very deep technical concepts (such as consistency in replication) that are hard to explain even to fellow engineers, let alone at a business meeting, but are easily countered by the competition's similar but vague claims.

We designed ScalienDB to be scalable in terms of data storage and processing. The intended use-case was Online Request Processing, ie. serving large number of requests behind a web application like a Software-as-a-Service product running in the cloud. This was also what our clients used ScalienDB for. ScalienDB was not meant to be used for batch processing like Hadoop. The cluster design assumes all nodes are in the same datacenter, where latency is small enough for synchronous replication. The design does not deal with geo-replication or geo-distributed data. From a programmatic point of view, our goal was to write a high-performance C++ server application with good operational characteristic that uses minimal memory and CPU resources.

In this paper we describe our experiences building and deploying ScalienDB at a handful of clients, enumerating interesting technical lessons learned. The intended audience is fellow systems programmers and implementors. We will not discuss the business experiences of running a startup, we will not give performance benchmarks and our goal is not to sell ScalienDB or convince the reader of its merits. ScalienDB continues to be open-source under the AGPL license \cite{AGPL} on Github, but it is no longer maintained.

\section{ Distributed Architecture and Data Model }
%%%%%%%%%%%%%%%%%%%%%%%%%%%%%%%%%%%%%%%%%%%%%%%%%%%

ScalienDB is separated into four main components: the controllers, which store the database schema and the cluster state, the shard servers which store actual data, the clients which use ScalienDB through the client library, and the web management console used for managing the schema and cluster state.

\textbf{Controllers.} The controllers form the controller quorum. A \textit{quorum} in ScalienDB is a set of servers which use replication to store the exact same data, similar to \textit{replication sets} in MongoDB. In the case of the controllers, every controller stores the exact same ScalienDB cluster configuration. The cluster configuration consists of the databases, tables, shard servers, and the quorum memberships, ie. which shard server belongs to which quorum. This cluster configuration data is very small in practice, less than 100KB in size. Also, since cluster configuration changes (eg. schema changes, new shard servers added) are very rare, the controllers experience very little load and are practically idle. The controller quorum can consist of just one controller (eg. a test cluster), or more (eg. the recommended production setting of 3 or 5). The controllers elect a \textit{master} controller using a distributed lease algorithm. The master makes all decisions, communicates with all other shard servers and clients. If the master controller dies, a new master is elected within seconds. \href{https://github.com/scalien/scaliendb/tree/master/src/Application/ConfigServer}{See code on Github.}

\textbf{Shard servers.} When a new shard server is added to the cluster, initially it does not store any data. It must first be added to a quorum, or a new quorum containing this shard server must be created. A ScalienDB cluster consists of one controller quorum and several shard server quorums. A quorum can contain any number of shard servers. As in the controller quorum, the shard servers inside their quorums store the exact same data using replication. The master controller appoints one of the shard servers in each quorum to be the \textit{primary}. Only the primary shard server accepts write requests from clients. Upon receiving a write request, the primary will initiate replication of the write command to the other shard servers in the quorum, and once it's complete it will respond to the client. A typical quorum consists of 3 shard servers. When one of the shard servers becomes unavailable, it will miss sending a heartbeat to the master controller, which will notice this and deactivate the shard server in the quorum. If this shard server was the primary, then a new primary will be appointed by the master. The whole process takes only a few seconds and does not cause problems for the client, since the client library explicitly handles such cases. The remaining two shard servers in the quorum will continue replicating. If and when the shard server comes back, it will perform catchup and the master will reactivate it in the quorum. \href{https://github.com/scalien/scaliendb/tree/master/src/Application/ShardServer}{See code on Github.}

Since both the controllers and shard servers in ScalienDB are replicated, this makes ScalienDB a highly-available, fault-tolerant distributed database. The single \textit{layer} of failure is the controller quorum, without it the cluster cannot operate, but there is no single \textit{point} of failure. Inside the (shard) quorums, any number of shard servers can fail since the controllers will just deactivate them. If all shard server in a quorum become unavailable, then all data (shards) that was stored there becomes unavailable, although other quorums will continue operating without problems.

\begin{figure}[htbp]
\begin{center}
\includegraphics[scale=0.5]{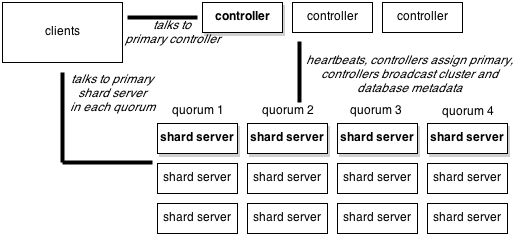}
\caption{ScalienDB cluster architecture.}
\label{default}
\end{center}
\end{figure}

\textbf{Sharding model.} Tables in ScalienDB are key-value namespaces. They are broken into 512MB continuous shards containing key-values. For example, a large table may be broken up into 3 shards by separating the key range like \texttt{[$\emptyset$, banana], [banana, tomato], [tomato, $\infty$]}. When a new table is created, it consists of just one shard. When this shard grows beyond the target size, it is broken in two, and so on (more details in the Storage Engine section). At this point both shards are in the same quorum, but now one of the shards can be moved into a different quorum. This makes ScalienDB scalable in terms of data storage and processing.

\textbf{Web management console.} Both controllers and shard servers include a simple HTTP server for checking  their status. Additionally, the controllers expose the cluster management API through a REST HTTP API. The web management console is a Javascript application which uses the API to manage the ScalienDB cluster. It lets the administrator create, delete and rename databases and tables, create new quorums, assign and remove shard servers  into quorums and manually reactivate shard servers in quorums. It uses simple color coding to signal the cluster state, for example when a shard server goes down and is deactivated in a quorum, that quorum is displayed in red on the management console. Although we felt that the management console is simplistic, it is still very friendly compared to what other NoSQL vendors offered and our clients loved it. \href{https://github.com/scalien/scaliendb/tree/master/webadmin}{See code on Github.}

\textbf{Clients.} The clients use the client library to connect to the ScalienDB cluster. The client library is explicitly designed to handle a large number of possible failure cases that can occur in a ScalienDB cluster, such as master failover, primary failover, quorum membership changes, etc. The goal, which we met successfully, was to avoid returning error codes in such cases to the application programmer. We felt that dealing with this is the job of the client library, after all ScalienDB is fault-tolerant and should hide the distributed aspects of the database if possible. Although this complicated the client library, we feel this was a good call. When there are unrecoverable problems in the cluster, our client library returns a structured error code to the client, consisting of a network status (were all request shipped?), a timeout status (the client library returned because of a timeout?) and a cluster error (eg. there is no master in the cluster). This proved to be a problem, because looking at these codes it was impossible to tell what was actually wrong in the cluster. This is a more general problem, because the client library only has a partial picture of the cluster state and what went wrong. \href{https://github.com/scalien/scaliendb/tree/master/src/Application/Client}{See code on Github.}

\textbf{Data model.} The data model is key-value based, with an additional table and database hierarchy familiar from the SQL world. A database contains tables, a table is namespace for key-value pairs. Supported key-value operations are (database and table arguments not shown):

\begin{enumerate*}
\item \texttt{Get(key)}: get the value.
\item \texttt{Set(key, value)}: set the value.
\item \texttt{Delete(key)}: delete the key-value.
\item \texttt{Truncate(table)}: delete all key-values in the table.
\item \texttt{List(start, end, prefix, num, direction)}: list all key-values between \texttt{start} and \texttt{end} that start with \texttt{prefix}, returning at most \texttt{num} keys. Both forward and backward listing directions are supported.
\item \texttt{Count(start, end, prefix, num, direction)}: return the count of key-values that a \texttt{List()} with the same parameters would return.
\item \texttt{Add(key, num)}: interpret the value belong to \texttt{key} as an unsigned 64 bit integer, increment it by \texttt{num} and return the value. Useful for generating IDs.
\item \texttt{Lock(key)}: attempt to acquire a time expiring lock (lease) for \texttt{key}.
\end{enumerate*}

This low-level functionality, plus the schema operations (not shown) such as \texttt{CreateTable()} are wrapped in language specific APIs. For example, the .NET API has a class \texttt{Sequence} which wraps \texttt{Add()} for generating unique IDs, which for efficiency increments the value on the server side by 1,000 and caches these values for the user, so only every 1,000 IDs requires a roundtrip to the server. Iterator classes wrap \texttt{List()} and also retrieve key-values in chunks using the \texttt{num} parameter for efficiency. The API also supports collecting write requests and sending them off in batches to save on disk writes and network roundtrips. Below is a sample C\# program:

\begin{verbatim}
string[] controllers = { "10.0.0.1:7080", "10.0.0.2:7080", "10.0.0.3:7080" };
Client client = new Client(controllers);
db = client.GetDatabase("testDatabase");
table = db.GetTable("testTable");
// batched sets
using (client.Begin())
{
    for (i = 0; i < 1000; i++)
       table.Set("foo" + i, "foo" + i);
}
// iterate
foreach(KeyValuePair<string, string> kv in
    table.GetKeyValueIterator(new StringRangeParams().Prefix("foo")))
        System.Console.WriteLine(kv.Key + " => " + kv.Value);
// truncate table
table.Truncate()
\end{verbatim}

\begin{figure}[htbp]
\begin{center}
\includegraphics[scale=0.5]{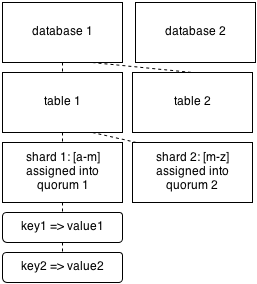}
\caption{ScalienDB data schema.}
\label{default}
\end{center}
\end{figure}

\textbf{Transactions.} ScalienDB supports light-weight transactions. Write operations (\texttt{Set}, \texttt{Delete} and \texttt{Add}) can be wrapped in transactions and they will be executed atomically, with the restriction that all writes inside a transaction must be limited to the same quorum. The isolation level of ScalienDB transactions is read committed. When the client is inside a transaction and issues a write request, the client locks those keys on the quorum primary and saves the write requests locally. The primary stores these locks in memory and expires them after a few seconds if the client does not commit the transaction, so these are really leases. This avoids lost locks in the case of client failure. When the user commits the transaction, the client library sends all saved up write requests including the commit to the server, where all locks are released after the writes completed. Read operations inside transactions see previous write commands, this is handled in the client library. A sample transaction using the C\# client library:

\begin{verbatim}
using (client.Transaction(quorum, majorKey)) // returns a Rollbacker guard
{
    table.Set("foo1", "bar1");
    table.Set("foo2", "bar2");
    client.CommitTransaction()
}
\end{verbatim}

The transactions feature is currently unfinished in ScalienDB. Our plan was to extend the data model using a \textit{major key}, so all operations that take a key should really take a major key and a key. The major key would be used to identify blocks of keys to never split such a block into two different shards. For example, in an application like Gmail the major key could be \texttt{userID}, this would make sure that all data belonging to a user is in the same shard and hence the same quorum, allowing per used transactions. The benefit is that transactions can be identified by major keys, the major key can be used to find the shard and thus the quorum that this transaction is limited to and all commands should be sent to.

\section{ Code }
%%%%%%%%%%%%%%%%

ScalienDB is roughly 100,000 lines of C++ code. It has no outside dependencies other than the standard filesystem APIs it uses. We chose to implement our own string, container and other utility classes, so the Standard Template Library is not used. One of the reasons we chose to reimplement these basic building blocks because it was a lot of fun. A more practical argument is efficient debug sessions: with all code being hand-written by ourselves, we could go deep on the call stack and still be looking at familiar code, which reduced bug fix times immensely. In retrospect we do not think this was a bad engineering decision, nor do we believe time spent implementing basic functionality did not pay off later in the debugging and bugfix stages.

Many of our coding decisions followed classic C++ design patterns such as ones found in books such C++ Gems and More C++ Gems \cite{CppGems,MoreCppGems}. We describe a few interesting pieces of C++ code.

\textbf{Intrusive data structures.} We used so-called intrusive data structures wherever possible. An intrusive data structure is one where the class being put in the container has to declare the member variables for the container functionality. The most common example in ScalienDB is the intrusive linked list \texttt{InList\textless T\textgreater}. For example, since we want to have a list of \texttt{TCPConnections}, \texttt{TCPConnection} includes \texttt{*prev} and \texttt{*next} pointers. This saves an extra \texttt{malloc()} and \texttt{free()} when inserting and removing items into the container, which is a good trade-off in a high-performance server application. Similar intrusive data structures are the intrusive stack \texttt{InStack\textless T\textgreater} and the intrusive red-black tree \texttt{InTreeMap\textless T\textgreater}. The use of intrusive containers was a major success. \href{https://github.com/scalien/scaliendb/blob/master/src/System/Containers}{See code on Github.}

\textbf{Separation of \texttt{Buffer} and \texttt{ReadBuffer} classes.} A \texttt{Buffer} is something which holds a piece of unstructed byte data, like a message received over the network or bytes read from the disk. It allocates and manages the memory, it can dynamically expand the backing memory area if new data is appended. It also features \texttt{printf()}-like functionality for formatted text output and \texttt{scanf()}-like functionality for formatted parsing. In a high-performance database we want to avoid making unnecessary copies of buffers. For example, when a message comes in over the network, it contains several messages wrapped into each other like a russian Matryoshka doll. We want to be able to parse and pass around the contained messages without making a copy of that part of the buffer or passing offsets. This is the problem solves by the class \texttt{ReadBuffer}. A \texttt{ReadBuffer} wraps a piece of memory, usually a \texttt{Buffer}, but has no facilities to allocate or free memory. It is basically a decorated \texttt{char*} pointer with a lot of useful functionality, like bounded formatted parsing functions. This separation of a "real buffer" and a "buffer pointer" has proved very useful in ScalienDB and probably also exists is many other server applications. \href{https://github.com/scalien/scaliendb/blob/master/src/System/Buffers}{See code on Github.}

\textbf{Server architecture.} ScalienDB is implemented as an asynchronous, event-based server. It uses \texttt{kqueue} on Darwin, \texttt{epoll} on Linux and Completion Ports (CP) on Windows to do network I/O. Since the very first version of this framework was written on a Mac, the \texttt{kqueue} model was used as the ScalienDB abstraction, and the other two platforms (Linux and Windows) were wrapped to emulate the \texttt{kqueue} model. This has proved to be quite a challange due to the differences in the APIs. With \texttt{kqueue}, we register for readyness events per socket, separately for read and write directions, and are notified edge triggered. For example, tell me when I can perform a non-blocking write of n bytes to this socket; when \texttt{kqueue} signals readyness, the program calls \texttt{write()}, and the operating system will take n bytes and send it off without blocking. On Linux, \texttt{epoll} is similar, but read and write signaling are handled together by the API. Since ScalienDB treats read and write separately, we have to emulate the \texttt{kqueue} model by saving read and write requests. For example, when ScalienDB wants to perform both read and write on a socket, \texttt{epoll} will signal separately for these, and we have re-issue the other, outstanding one, so we have to keep track of these outstanding I/O operations. Windows' CP API uses a truly asynchronous model, where completion of the \texttt{read()} or \texttt{write()} is notified on the CP when it's finished. We hit the biggest problem on Windows trying to cancel outstanding I/O operations: the ScalienDB application assumes that it can cancel outstanding I/O operations at any time synchronously, which is trivial in the other models, but with CP cancel is also an asynchronous operation whose completion is signaled later. Hence, on Windows, we were forced to make copies of all buffers associated with the I/O operation and pass those to the operating system, because in case of a cancel request the operating system may use these buffers until the cancel succeeds, but ScalienDB assumes that the cancel executed instantaneously and then potentially wants to use the buffer. In retrospect, given that we have to support all three operating system's model, using the \texttt{kqueue} model is not a good choice for the application level abstraction. We believe the Windows model is the most general, even though managing reads and writes in the \texttt{epoll} model would still be required. \href{https://github.com/scalien/scaliendb/tree/master/src/System/IO}{See code on Github.}

\textbf{Client library.} The ScalienDB client library is written in C++ and uses the same asynchronous I/O code as the server, running in a separate thread. The client library takes requests like \texttt{Get} and passes it on to the main network thread and blocks waiting. The network thread performs the network I/O on behalf of the client and then wakes up the client. From the perspective of the application programmer the ScalienDB client is a regular, blocking library. The client is implemented as a state machine, where the client requests and the events it receives from the networking thread trigger the state transitions. For example, if the primary node in a quorum goes down, it will get deactivated, and the client library will receive a new cluster configuration state from the controllers. Upon examining this, it will notice that any write requests sent out to the now deactivated node will have to be resent to the new primary. The ScalienDB client library is available for Python, Java and .NET. We use the Simplified Wrapper and Interface Generator (SWIG) to wrap our C++ API and expose it to the mentioned languages. Additional language specific code thinly wraps the SWIG wrapper to make the API conform to each language's idioms. In retrospect, the use of SWIG caused a lot of problems because debugging the client library was very problematic. Clients would report problems when using the Java or .NET client, but debugging the underlying C++ code was not possible. In retrospect we would rather implement the client library for each programming language and use each environment's API to perform network I/O. \href{https://github.com/scalien/scaliendb/tree/master/src/Application/Client}{See code on Github.}

\section{ Replication }
%%%%%%%%%%%%%%%%%%%%%%%

ScalienDB uses the Paxos algorithm \cite{PaxosMadeSimple} for replication, both the controllers for cluster meta data and the shard servers for shard data. In ScalienDB, nodes are organized into quorums. A node can only take part in replication once it has been assigned to a quorum. A node can be part of more than one quorum, in this case it receives writes from more than one quorum, but this feature was never used in production by our clients, so it could be removed from the code to make the architecture simpler.

\textbf{Paxos and other replication schemes.} Paxos is a distributed algorithm for reaching consensus in the presence of faults (message loss, message reordering, duplication, arbitrary transit times, node failure). ScalienDB uses Paxos for determining what the next database command should be. The nodes in a quorum run a round of Paxos and reach consensus on the next database command, then all of them execute the command on their local copy of the database. Paxos is consistent, which means that there is no possibility of conflicts or divergent database versions: the majority of replicas in a quorum always have the same exact local copy of the database, with a minority possibly trailing behind but never diverging.

In the currently popular set of NoSQL databases, three types of replication schemes are widespread: (1) lossy replication (2) eventually consistent replication and (3) consistent replication (Paxos).

\begin{enumerate*}
\item Lossy replication is when a database command can be completely undone even after it has been acknowledged to the end-user. For example, in MongoDB it's easy to see such cases when the database is configured to write to disk to the master node, but replication is asynchronous to the slaves. If one writes continuously to the database, then kills the master, eventually the slaves will elect a new master. Writes that have not yet propagated to the slaves before the old master was shut down (but acknowledged to the end-user) will be lost in this case: the old master will come back, find the last common point in its log, and throw away its local unpropagated writes (they are actually saved to a file, it's up to the administrator to examine them by hand).
\item Eventually consistent replication accepts that there may be divergent version of data in the distributed database due to failures or concurrent writers, and uses vector clocks to keep track of these divergent versions (eg. Riak, Cassandra). The benefit of eventually consistent replication is that the system can take writes when a majority of nodes are down, and that replication doesn't have to be synchronous, in the sense that the end-user doesn't have to wait for the nodes to internally run a replication algorithm like Paxos. The cost of this is that there may be divergent versions of the same data on different nodes, due to concurrent writers or failures. In this case, conflict resolution occurs, which tries to reconcile the different versions of data and results in a new, resolved version of the data. It is worth noting that some vendors claim that when using their eventually consistent databases such that \texttt{W+R\textgreater N} \footnote{ \texttt{N} is the number of nodes storing a replica of the data, \texttt{W} is the number of nodes a write has to reach before it is acknowledged, \texttt{R} is the number of nodes that have to be examined before returning a \texttt{Get} result.} "strong consistency" is achieved, but this is misleading. First, in many of these databases, if a node becomes unavailable, another node not previously taking part in the replication of the data joins, which means that the definition of \texttt{N} is problematic because membership in replication is dynamic. Second, putting aside such scenarios, all \texttt{W+R\textgreater N} guarantees is that the reader will "see" all previous writes to the data, but not that there will be no conflicts. For example, if \texttt{W=1} and \texttt{R=N}, the inequality trivially holds, and with \texttt{N} concurrent writers there can be \texttt{N} divergent version of data in the system, which will undergo conflict resolution when a reader comes along and sees all \texttt{N} versions. The classic example of an eventually consistent database is Amazon's Dynamo \cite{Dynamo}.
\item Paxos is an algorithm for consistent replication. The algorithm trades some availability for consistency: a majority of nodes have to be up and communicating for the distributed algorithm to make progress. In return, Paxos guarantees that the local copies of the database will never diverge: all local copies of the database execute the same database commands in the same order. The algorithm itself is used for replicating these database commands. It's easy to see why such an algorithm requires a majority: if it did not, then two minorities could run the algorithm and execute different database commands, leading to divergent distributed states. Majority guarantees that any two sets of nodes doing something will have at least one node in common, and the behaviour of this common node guarantees consistency. Paxos has two notable characteristics. First is that it is a synchronous replication algorithm, in the sense that when a user sends a database write command to a node, that node must run a round of Paxos with the other nodes before it can acknowledge the command. Second, the algorithm requires the nodes to write their state to disk, so they will be in the same state if they are restarted and are able to keep the promises they made while running the algorithm. The upside is that a distributed database using Paxos has the appearance of a single-node database to the end-user, because it hides the distributed nature of the underlying fabric, as no inconsistencies can occur at the cost of requiring a majority of nodes to be up.
\end{enumerate*}

\textbf{A concise description of Paxos.} There are three roles in Paxos: proposers, acceptors and learners. In the abstract definition of Paxos, these are different nodes, so there is a set of proposers, a set of acceptors and a set of learners. In ScalienDB, all nodes act as all three. Proposers receive write commands from end-users and propose these commands to be the next command to be executed, this is called the proposed (and later accepted, learned) value. Acceptors receive and reply to messages from the proposers, otherwise they are passive.  They make promises to the proposers and write their state to disk so they can keep their promises if they are restarted. Once a round of replication completes, the proposer sends out learn messages to the learners, which passes it on to the local database for execution. In ScalienDB terminology, the three phases of Paxos are (1) the Prepare phase, (2) the Propose phase and (3) the Learn phase. In the prepare phase the proposer and the acceptors co-operate to prepare for the propose phase. In the propose phase the proposer and the acceptors co-operate to reach consensus on the accepted value. In the Learn phase the proposer sends the accepted value to the learners.

\begin{enumerate*}

\item \textbf{Proposer's algorithm.} The proposer wants to propose a value. It send out PrepareRequest message with a unique \texttt{proposalID} to all the acceptors. The \texttt{proposalID} must be unique, so the proposer uses its \texttt{nodeID} and \texttt{runID} which is incremented every time the node reboots (stored on disk), plus an in-memory counter. The PrepareRequest message does not include the proposed value.

\textbf{Acceptor's algorithm.} Acceptors store a \texttt{promisedProposalID}, which is initially $\emptyset$. Upon receiving a PrepareRequest message, it checks whether $\texttt{msg.proposalID} < \texttt{promisedProposalID}$, in this case it sends a reject message. If $\texttt{msg.proposalID} \geq \texttt{ promisedProposalID}$, it sets \texttt{promisedProposalID := msg.proposalID} and writes to disk. In other words, it promises to the proposer that it will not cooperate with other proposers sending PrepareRequest messages with a lower \texttt{proposalID}. It then sends a PrepareResponse, which contains the acceptor's currently \texttt{acceptedProposalID} and \texttt{acceptedValue}, or $\emptyset$ if it has not accepted a proposal previously.

\item  \textbf{Proposer's algorithm.} If it received PrepareResponses from a majority of acceptors, the proposer can advance to the Propose phase. If it received PrepareResponses that are all $\emptyset$, the proposer is free to propose its own value in the propose phase. If not, it must propose the value contained in the PrepareResponse with the highest \texttt{acceptedProposalID}. This is the part of the algorithm that guarantees that previously accepted values are not undone or overwritten by subsequent runs of the distributed algorithm. The proposer sends out ProposeRequests to all acceptors containing the \texttt{proposalID} and the \texttt{proposedValue}.

\textbf{Acceptor's algorithm.} Upon receiving a ProposeRequests, it again checks whether $\texttt{msg.proposalID} < \texttt{ promisedProposalID}$. If not, then it sets \texttt{acceptedProposalID := msg.proposalID} and \texttt{acceptedValue := msg.value} and writes to disk. It then sends a ProposeResponse message back to the proposer, or else a ProposeReject message.

\item  \textbf{Proposer's Algorithm.} If it receives ProposeResponse messages from a majority of acceptors, then it was successful. It sends out Learn messages containing the \texttt{proposedValue} to all learners.

\textbf{Learner's algorithm.} Upon receiving a Learn message, the learner takes the value, parses it as database commands and executes it against the local database.

\end{enumerate*}

\href{https://github.com/scalien/scaliendb/tree/master/src/Framework/Replication/Paxos}{See code on Github.}

\begin{figure}[htbp]
\begin{center}
\includegraphics[scale=0.5]{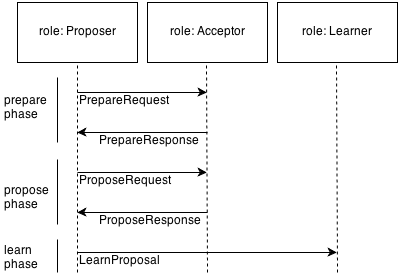}
\caption{A round of Paxos.}
\label{default}
\end{center}
\end{figure}

Once a value is learnt, all subsequent runs of the distributed algorithm, even by different proposers, will yield the same result. Leslie Lamport, the inventor of Paxos proved that Paxos is the minimal replication algorithm which works in the presence of faults and guarantees consistency. If a majority of nodes are able to communicate one of the proposers will eventually succeed, if it uses appropriate timers to restart in case of concurrent proposers or failures. There is no possibility of a static deadlock in the sense that proposers can't lock each other out. Dynamic deadlock such as concurrent proposers are always possible in distributed system, but this is easy to work around with timeouts.

\textbf{Paxos in ScalienDB.} We use Paxos with a few optimizations and improvements. The Paxos framework is programmed such that the subset of nodes required for progress is configurable. The controllers use majorities, as in the original definition. The shard servers use what we call TotalPaxos, where we require all nodes to take part in the algorithm. It is easy to see that this does not break the algorithm's properties, but it does increase the requirements on the availability of the nodes. We do this because we use a different mechanism to allow for node failures: if a shard server goes down, the controllers will notice that it's not sending heartbeats anymore, and they will deactivate the shard server in the quorum, so that replication can continue. This way the controllers can deactivate $n - 1$ nodes in an n-way replicated quorum, which is the way ScalienDB beats the so-called CAP theorem. Because of the use TotalPaxos, the controllers can assume that nodes that are part of a quorum are all up-to-date, which is necessary when removing and adding nodes to a cluster to not break the consistency guarantees of the distributed system. \href{https://github.com/scalien/scaliendb/tree/master/src/Framework/Replication/Quorums}{See code on Github.}

Another modification is the use of PaxosLease \cite{PaxosLease} to elect a master node in the controller quorum. PaxosLease is variation of Paxos invented at Scalien for electing a master node in a consistent manner, without deadlocks. It uses the same basic idea as Paxos, with the difference that the proposed value is always the request of a proposer to receive the master lease for some time. When an acceptor accepts a value, it starts a timer for the duration of the proposer lease, and honors its promise not to help another node until the timeout occurs. When it occurs, it clears it state, and another proposer can become the master node, unless the existing master has extended its lease. Since a lease is by definition ephemeral, in PaxosLease the acceptors don't have to write their state to disk, yielding a very efficient and elegant algorithm. The shard servers also use leader-based Paxos, but instead of using PaxosLease to elect a primary, the master controller simply appoints a primary in each quorum. \href{https://github.com/scalien/scaliendb/tree/master/src/Framework/Replication/PaxosLease}{See code on Github.}

The presence of a leader (master for controllers, primaries for shard servers) in Paxos allows for a very important modification in Paxos, invented by Google for use in its Chubby database \cite{Chubby}, called MultiPaxos in ScalienDB terminology. In regular Paxos, the acceptors must write to disk two times per round, once in the Prepare and once in the Propose phase. But, if there is only one proposer who proposes values, the leader can skip the Prepare phase and jump straight to the Propose phase. The idea is that we can always assume that the leader sent out a Prepare with \texttt{proposalID := 0} and that all acceptors accepted it, as long as there is only one proposer who makes this assumption (the leader). Basically, with this modification Paxos turns into a simple case of the proposer sending out its values to all acceptors. One caveat is that in the case of leader failover, ie. when another node was the leader in the current round of Paxos, we must account for the fact that the acceptors may be in some state due to the previous proposer, so we cannot make this assumption and skip the Prepare phase. For this reason, when a node becomes the leader, in the first round of replication it always runs a full round of Paxos to clean out the acceptors.

Another trick is commit chaining. When running Paxos with leaders, the number of disk commmits and roundtrips is one (in the Propose phase). However, it is in the Learn phase that the learner actually executes the database commands against the local copy of the database, which requires another disk commit. Commit chaining is a trick whereby the database is only commited in the Propose phase, but not in the Learn phase when the database commands are executed. We simply wait for the next Propose phase for the commit. One apparent problem with this is if the node restarts after a Learn phase but before the next commit. This is actually not a problem, because in this case the node, when it comes back, will be (at least) one Paxos round behind, so the proposer will either re-run that previous round of Paxos (which must yield the same result), or if the other nodes have meanwhile moved on they will simply send this node that round's Learn value in the process of catchup.

A related optimization is possible in the Learn phase to save network bandwidth. All nodes are both acceptors and learners, and acceptors already store the accepted value which is almost always the same as the learned value. The proposer by default sends a learn message that just contains the \texttt{proposalID}, and the learner checks whether it matches its acceptor's \texttt{acceptedProposalID}, in which case it takes its stored \texttt{acceptedValue} as the learned value, saving network bandwidth. If it does not match, it sends a request to the current leader to send the learned value. This can only happen during leader failover or node failure, not during normal steady-state operation.

Paxos is an algorithm for reaching consensus on one value. In ScalienDB, we pack several database commands into each proposed value, up to a few MB in size. Nevertheless, we must run subsequent rounds of Paxos to replicate database commands that arrived since the last round was replicated. Each round is identified by a \texttt{paxosID}, which starts at 1 for each quorum and increases indefinitely. The exact semantics of running subsequent rounds of Paxos are beyond the scope of the algorithm itself, and are ScalienDB-specific. For example, Chubby allows several rounds of Paxos to be run in parallel, which has the advantage that a small (few KB) round doesn't have to wait for a previous large (few MB) round to complete \cite{Chubby}. However, we felt this would further complicate the implementation, so ScalienDB does not do this; rounds of Paxos are run strictly one after the other. \href{https://github.com/scalien/scaliendb/tree/master/src/Framework/Replication/ReplicatedLog}{See code on Github.}

The controllers use standard majority Paxos, while the shard servers use TotalPaxos. In both cases, it is possible for nodes to be left behind in replication. For the controllers, if a node goes offline, if a majority is still available, they will continue replicating. For the shard servers, if a node goes offline, the controllers will remove it from the quorum (unless it's the last one) and the remaining nodes will continue to replicate. In both cases, when the node comes back it will usually have missed several rounds of replication, so it must catch up to the rest of the nodes. The controllers store meta data about the cluster and the database schema, which is very small, usually a few tens of KBs in size. So in the case of controllers, the lagging node simply copies the current state, sets its \texttt{paxosID} to the quorum's \texttt{paxosID}, and it can resume participating in the current round of replication. In the case of shard servers, catchup is a much more difficult process, since they potentially store hundreds of GB of data, so always copying over the entire database is not optimal. Hence the shard servers store (by default) 20GB of the replicated log on disk, and when a node is lagging, they will help it catch up by sending it these rounds as Learn messages. This has interesting implementation issues: from which node should the lagging node catch up? what if there are several lagging nodes? what is the speed of catchup compared to the normal replication speed? In ScalienDB, all lagging nodes catch up from the primary node in the quorum, mostly because not much can be gained from using another node, as all nodes experience the same disk load during normal operation, and catchup is disk intensive, but not memory/CPU intensive. The speed of catchup is production proved to be about 50-100\% faster than normal replication, which is necessary for replication to finish at some point. However, this is not by design, it's by accident, probably related to less roundtrips and less in-memory allocation overhead.

What happens when a node has been offline for so long that the 20GB of the replicated log doesn't go back long enough? In this case, the node has to copy over the entire database from another node in the quorum, this is called database catchup in ScalienDB. As in the log based catchup path, copying happens from the primary node. In production, it turned out that this code path in ScalienDB has some inefficiencies, because even with enterprise-grade disks the copying maxed out at about 10-15MB/s, which was deemed unacceptable by the client's infrastructure engineers. As an alternative, it is possible to use OS-level file copy to copy over the database, which can proceed at much quicker speeds (e.g. 50-100MB/s). This is what our clients ended up doing. Other NoSQL databases like MongoDB don't have a built-in database catchup code path at all and the administrator has to do this by default. This is a problem that could probably be fixed in the code, but we never got around to it.

A related topic to catchup is activation. This is what happens when a shard server has caught up and can be reactivated in the quorum. There is no activation in the controller quorum, because controllers never get deactivated, only shard servers in the shard quorums. Activation seems simple, all the controller has to do is tell the primary shard server to treat the node undergoing activation as part of the quorum. But it's not that simple, for example the primary is continually replicating new rounds, the lagging node is continually catching up, at what point can activation actually occur; the controller only has delayed state information about them. Or, the primary may at the point of activation be at some phase of running a round of Paxos, and introducing a new node could break the algorithm. So the primary always restarts the current round when activation is going on and runs a full round of Paxos. In practice, this process of activation proved to be a problematic part in ScalienDB, with many hard to reproduce and hard to understand bugs arising from it due to the distributed nature of the problem and many processes interacting (catchup, Paxos, Paxos optimizations). Although we fixed all known bugs in ScalienDB related to activation, it would be nice to see a formally specified and proven algorithm for solving this problem.

\section{ Storage Engine }
%%%%%%%%%%%%%%%%%%%%%%%%%%

\textbf{Why we abandoned BerkeleyDB.} When building Keyspace, we wanted to avoid writing our own disk storage engine, as we wanted to concentrate on replication. We chose BerkeleyDB, because it promised to be an industrial strength plug and play storage engine. BerkeleyDB has several storage engines to choose from: Data Store (DS), Concurrent Data Store (CDS) and Transactional Data Store (TDS). Because of the semantics required by our replication algorithm and because we wanted strong write guarantees we had to use TDS. TDS is theoretically the safest of the three as it uses a redo log to protect against database corruption if the process is terminated during mutation of the database file. However, in our experience BerkeleyDB proved to be unreliable for production use, as we constantly ran into show-stopper bugs and data loss:

\begin{enumerate*}
\item BerkeleyDB frequently corrupted its database files (with TDS, transaction logs, correct flags). This seemed to be related to open cursors when the program shut down. Searching Google revealed many similar issues reported by users of open-source programs which used BerkeleyDB underneath.
\item A few GBs of transaction log files took hours to replay upon each start, orders of magnitude more than the disk read speed, which is basically temporal data loss because the database takes so long to come back up. Also, BerkeleyDB offers no progress report on what it's doing.
\item What we called the \textit{long put problem} during heavy write tests: on some platforms (Darwin, Linux EC2), sometimes \texttt{put()} operations would take arbitrary long (eg. more than 5 seconds) to execute, even with small overall database size (100MB). We could not reproduce this problem on physical Linux servers.
\item Database checkpointing (which removes the redo logs) effectively blocks the database.
\item No API feature to return the top element of the underlying b-tree to get an approximation of the middle key, which we needed in order to support splitting shards into two parts for re-balancing.
\item No API to iterate the keys quickly, at the speed of the underlying disk in non-sorted order. BerkeleyDB only allows iteration in sorted order, which means jumping around on disk per the b-tree's pages, which limits cursor throughput to a few MB/s (depending on page size), making it unusable to use a cursor to copy the database to another server during catchup.
\item Hotbackup sometimes took more than 24 hours for a 10GB database.
\item BerkeleyDB TDS was very slow for databases \textgreater 100GB.
\item Large number of programmatic options for setting up and tuning the database leading to combinatoric blowup of the configuration space. Clashing options were silently ignored, possibly leading to unexpected semantics with reduced consistency guarantees. Almost every database operation such as put, get and iterators also takes flags, making it hard to understand what the expected semantics and performance trade-offs are, and under what circumstances the TDS engine actually does not corrupt data in case of program restart or crash, if any.
\end{enumerate*}

Due to the particular license Oracle ships BerkeleyDB under, we were not allowed to distribute Keyspace with the BerkeleyDB source code that we developed against. This means BerkeleyDB was an outside library dependency of Keyspace. Our users had to separately download and install BerkeleyDB to get Keyspace to run, or if they were fortunate a good version of BerkeleyDB was pre-installed on their Linux machines. We felt this step turned many users away, either up front or when they ran into issues trying to download and compile BerkeleyDB and Keyspace. Although BerkeleyDB is a fairly old and established product, its API still changes from version to version (eg. renamed flag constants), so if a user's system had a BerkeleyDB version a few versions behind, she would get nasty compile-time errors. Almost all of our users ran into issues with BerkeleyDB trying to get Keyspace to run.

Finally, the dependence on an Oracle product whose licensing might change put us in an uneasy business situation. Additionally, when negotiating commercial licensing with potential users, explaining the relationship of Keyspace to BerkeleyDB, and the related licensing issues put an additional spin on the already difficult and time consuming process of enterprise sales.

\textbf{Writing our own storage engine for ScalienDB.} When starting development work on ScalienDB, after having spent an inordinate amount of time trying to work around the oddities of BerkeleyDB, one of the major design decisions was to abandon it and write our own storage engine. After examining the available open-source storage engine libraries, none seemed up to the task of being a general purpose, transactional data store to use in a real database with the particular requirements and optimization possibilities of ScalienDB's architecture. The bad experiences with BerkeleyDB also factored into this decision. We felt writing our own high-speed storage engine would not take too long and would be a good overall trade-off:

\begin{enumerate*}
\item Optimization possibilities 100\% in our hands.
\item Easy to maintain, debug and optimize because it's our own code.
\item No outside library dependencies for easy one-step installation.
\item We own the code, no complicated licensing issues.
\end{enumerate*}

In retrospect, we think this was a good call. Today, a few more options are available, such as LevelDB by Google, which is very similar to the ScalienDB storage engine, as both are based on Bigtable \cite{Bigtable}. However, the storage engine is such a central module in a database system that treating it as a black box is not a good trade-off in the long term, so we would still opt to write our own storage engine today. Unlike BerkeleyDB, the ScalienDB storage engine is not available as a stand-alone library, it cannot be used separately.

We designed the storage engine to address the shortcomings of the BerkeleyDB engine discussed above:

\begin{enumerate*}
\item Unlike BerkeleyDB, it is safe, because it uses a redo log and the database is never corrupted.
\item Playing back the redo log is fast, comparable to the linear read speed of the disk.
\item Write speed is independent of database size.
\item Read speed is linear in the database size (worse than b-trees' logarithmic dependence).
\item Supports fast iteration of keys in disk order.
\item Can return the approximate middle key of a shard of data at constant cost.
\end{enumerate*}

The engine is optimized for fast writes, writing to disk is linear because no file is ever changed or rewritten. Once a file is written, it is never changed, only deleted later once it's redundant because its contents have been written to a newer file. This is true for log files, chunk files and the table of contents files.

The architecture is centered around chunk files. As writes come in to the storage engine, they are written to the redo log and collected in in-memory chunk files. Once an in-memory chunk file reaches a certain size (default 64MB), it is written to disk. The keys within a chunk file are sorted, but chunk files are not sorted relative to each other, they contain values as they came in from the user. This means that when performing a \texttt{Get} request and looking for a specific key, potentially a large number of chunks have to be examined. To speed this up, we use bloom filters. A bloom filter is a probabilistic data structure which can, at the cost of a hash computation return:

\begin{enumerate*}
\item \texttt{NOT\_PRESENT}: they key is not stored in the chunk file.
\item \texttt{MAYBE\_PRESENT}: it's possible that the key is stored in the chunk file.
\end{enumerate*}

Bloom filters work by computing a hash for each key stored in the chunk file and OR'ing these, and we store this bitwise OR in each chunk file on the bloom page. When looking for a key, the key is hashed and the bits are examined on the bloom page. If at least one is not set, then we can be certain that the key is not stored in the chunk file (\texttt{NOT\_PRESENT}). If all the bits are set, then we have to look inside the chunk file, as this key may be present (\texttt{MAYBE\_PRESENT}). The goodness of this data structure depends on how many bits we use for the bloom filter to reduce hash collisions and the likelyhood of a false positive. The storage engine uses bloom pages sized so that the probability of false positives is 10\%. \href{ https://github.com/scalien/scaliendb/blob/master/src/Framework/Storage/BloomFilter.cpp}{See code on Github.}

When a chunk file is written to disk, it contains a header page, an index page, a bloom page and several data pages each 64KB in size of sorted key-values. The header page stores information like the smallest, largest and middle key stored in the chunk file. The index page contains the first key and offset of each data page. The bloom page contains the bloom filter. The data pages can contain two types of entries: \texttt{Set(key, value)} and \texttt{Delete(key)}.

It is interesting to note that in the first version of the storage engine, parsing the data pages proved to be a bottleneck. In this first version, the key-values on a data page were parsed into a red-black tree in memory. However, constructing this tree, which includes \texttt{malloc()} calls for allocating the nodes and buffers for the tree proved to be an unexpected CPU hog, and also caused memory fragmentation on 64-bit Windows platforms. As an optimization, we settled on reading the keys and values into a large buffer and using binary search.

Keys can be deleted, but chunk files storing the previous key are never changed. Hence, \texttt{Delete} operations are stored just as \texttt{Set} operations. Relative to each other chunk files contain operations in chronological order, and they must be examined in this chronological order. When executing a \texttt{Get} request, if the first entry found is a \texttt{Delete}, then \texttt{NOT\_FOUND} is returned.

This scheme can lead to data duplication: old key-value pairs are no longer relevant if the key has been deleted or a new value set in a more recent chunk. To clean up old chunks, every once in a while the storage engine merges chunks in the background. Chunk merging proceeds by taking all the chunks and iterating all key-values in parallel. If a key-value is only found in one of the chunks, it is emitted and written to the merged chunk. If a key-value is found in several chunks, the one from the more recent one is used. If that is a \texttt{Delete}, then it can be skipped and nothing emitted altogether, since there will be no chunks behind the merged chunk. After chunk merging is complete, the old, merged chunks can be safely deleted off the disk. Since chunk merging involves the linear reading of several chunk files we use pre-reading to optimize. At first we feared that chunk merging would be an expensive operation that could prove to be a bottleneck in our architecture, however in real-life production workloads with enterprise grade hard disks chunk merging proved to be very quick relative to the overall ability of the database to take writes and did not cause any performance problems.

ScalienDB was designed to be a scalable database. This means that when new nodes are added to the cluster, we want to be able to re-balance the data. The approach ScalienDB takes is to pre-shard tables into shards of equal sizes. A shard is a piece of a table defined by a first and a last key: it stores all key-values between the first and last keys. Shards are then the unit of data scalability: when re-balancing, shards are moved over from old nodes to new nodes. More precisely, shards are moved from one quorum to another. The overall hierarchy of the storage engine is depicted in Figure 4.

\begin{figure}[htbp]
\begin{center}
\includegraphics[scale=0.5]{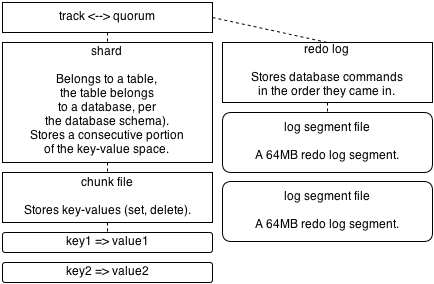}
\caption{Storage engine architecture.}
\label{default}
\end{center}
\end{figure}

A track is the highest level in the storage engine hierarchy. ScalienDB creates one track for each quorum it is in, as each quorum corresponds to a separate instance of Paxos, which needs to be commited separately, and each quorum stores separate data.

When a table is first created, it contains just one shard. As data is written to it, several chunks are created. Since the chunks may contain duplicate data relative to each other, the exact logical size (the size of the data if all the chunks would be merged) of the shard is not precisely known at all times and the physical on-disk size is used. When a shard reaches a certain size, it is split in two using the approximate middle key. The middle key is approximated using the middle keys stored in the chunk files: the middle keys are sorted and the middle element is taken. Shard splitting is in itself a purely logical operation: a new shard is created, which inherits all the chunk files of the old shard (a chunk can belong to more than one shard in ScalienDB). Eventually chunk merging happens on both shards, which takes into account the first and last key of the shard, and the keys outside the shards are discarded when the new chunks for the shards are written. It is possible in this scheme for shard splitting to occur at a poorly chosen middle key. This is especially true is one of the chunk files is much larger than the others, which occurs after merging: the merged chunk will be large, newly written chunks will be of a standard size. This is counterbalanced by aggressively merging chunks so most shards contain just one large chunk. In ScalienDB the default shard size was 512MB, the default chunk size was 64MB.

An interesting problem encountered was related to the redo log size. When ScalienDB is restarted, it reads through the redo log and re-executes all operations that have not been written to file chunks in the previous run; this is called recovery. The length recovery takes is determined by the size of the redo log, so ideally we'd like to keep it small. In ScalienDB, the redo log was written to log segments, each 64MB by default, and the overall redo log size is 20GB by default. A log segment can be deleted from disk if all data contained in it has been written to chunk files. In-memory chunks of shards that are receiving a lot of writes will often reach the 64MB limit and be written to disk, however it is enough if a log segment contains just one write that belongs to a chunk that is still in-memory: in this case the log segment cannot be deleted. If the 20GB limit is reached, ScalienDB will force these chunks to be written to disk, even if they are smaller than 64MB, so the tail of the redo log can be deleted. This means that certain shards, such as ones belonging to tables storing meta information that receive little writes will contain lots of small chunks, a few 100KB in size each that were forced to disk by this mechanism. On the other hand, if there are a lot of shards, they will all be competing for the same redo log space, and the average file chunk size will decrease. For example, if there are 1000 shards, the average chunk size will decrease to 20MB (=20GB/1000), even though the hard limit is 64MB. These issues were the subject of major concerns on our part. For example, one client ran ScalienDB with chunk merging completely turned off for a month on a 100-1000GB database, which resulted in tens of thousands of chunk files of varying sizes. When the client finally turned chunk merging back on, all chunks were merged very quickly in a matter of hours, so these issues were --- luckily --- not a concern on enterprise grade production hardware. Nevertheless, chunk merging was one aspect of our design and code that we were always worried about and kept going back to: what logical conditions should trigger chunk merging (number and size of chunks), when should we pause chunk merging to give other, higher priority logic additional resources. \href{https://github.com/scalien/scaliendb/blob/master/src/Framework/Storage/StorageChunkMerger.cpp}{See code on Github.}

For example, ScalienDB will dynamically pause chunk merging when processing lots of reads that are hitting the disk. Also, it will not perform merges if there are concurrent iterators open on the database. We also noticed that we had starvation type bugs, where some shards were never merged because we always merged the first shard we found to be a good candidate; this was later changed so the largest shard is always merged. We also realized that there are two distinct conditions when a chunks need to be merged: (i) the shard is a result of a split, so the chunk files contain data that is outside the shard, and (ii) the shard has a lot of chunks. We eventually created separate conditions for the two cases, and merge aggressively in the first case, but only if there are more than 10 chunks in the second case. This proved to be a good trade-off in production. \href{https://github.com/scalien/scaliendb/blob/master/src/Framework/Storage/StorageEnvironment.cpp}{See code on Github.}

A ScalienDB-specific optimization example in the storage engine is the use log type shards for the replicated log. ScalienDB stores the results of each round of Paxos in the database in a special system database not visible to the user. This data is of the form \texttt{replication round number $\rightarrow$ learned value}. Since this data is stored in chunks the way user data is, the optimization possibility here is that these chunks never have to be merged, because this value is never overwritten thanks to Paxos semantics. Eventually it is deleted, because we cannot afford to keep an infinite replicated log on disk. The size of the replicated log stored on disk (to help lagging or new nodes to catch up) is 10GB by default. Log entries more behind the head must be deleted. This is also performed in an optimized way, instead of issuing a \texttt{Delete} command for those keys, the way we had to do it with BerkeleyDB, we simply delete those chunk files from disk.

The storage engine uses a common file system abstraction that hides the OS (Posix and Windows) specific filesystem API. \href{https://github.com/scalien/scaliendb/blob/master/src/System/FileSystem.cpp}{See code on Github.} Since there is no good asynchronous filesystem API on either platform, we use synchronous filesystem calls to create a common, synchronous filesystem API. However, disk access has to be asynchronous, since a database cannot afford to block waiting for disk access. We chose to implement this in the storage engine itself: whenever we perform reads or writes, we send this request over to a different thread, and then receive the result in the form of an event. At first some of the calls were synchronous, and we converted more and more calls to be asynchronous over time, which led to some ugly code in places, especially where a number of reads have to be performed in succession, such as \texttt{Get}s looking through several chunk files, or iterators. Other parts, such as writing chunks to disk or merging chunks were not affected since these jobs are completely asynchronous anyway. Currently, the only filesystem access that is synchronous is the writing of the table of contents file to disk. This file contains the database metadata such as databases, tables, shards and the shard $\rightarrow$ chunk mapping. It is only a few KB is size, however the write needs to be synced, and syncing can be an expensive operation, so this can in theory lead to noticable blocking behaviour. This is problematic because if the node blocks, it can lose the primary lease, miss a few rounds of replication and become deactivated in the cluster. These are all occurences that are handled by ScalienDB, however operations teams still alert on these occurences and demand that they do not occur in normal operation when no real fault occured. This last synchronous disk write would also have been converted to be asynchronous with time.

At about 10,000 lines of code the storage engine is only about 10\% of the overall ScalienDB source code, however it is the most complex and error-prone part. It is the one part of the code base where, over time, we broke most of the rules of good object-oriented design and e.g. have large classes containing too much logic, classes with all public members, a mix of functions which run in the main thread versus asynchronous, differing return semantics for asynchronous calls, etc. First, most of the other parts of the ScalienDB code were a rewrite of our previous tech, Keyspace, so we had previous experience and were able to design the software a priori. The storage engine however was our first storage engine ever, and as such carries all the signs of experimentation and learning relative to the rest of the codebase. Second, the storage engine carries a lot of intrinsic complexity and optimization: writing and recovering the redo log, chunks and their associated data structures such as header, index, bloom and data pages, serializing chunks, writing chunks to disk, merging chunks, shards, shard splitting, page caching, iterating data using cursors, a lot of asynchronous operations, and of course the burden of logical correctness of all this, which if broken potentially leads to data loss. This was a major limiter in our willingness to refactor otherwise working code, esp. once the code was in production at clients.

\section{ Discussion }
%%%%%%%%%%%%%%%%%%%%%%

ScalienDB ran in production at a handful clients for about a year, but it never saw a full-blown release to a representative number of clients. Even at this small number of sites, in 2011-2012 when ScalienDB was deployed, a fair number of serious bugs were encountered. In our experience this is normal for systems software where there was simply not enough human or hardware resource available to do proper, large-scale testing. We never had more than 2 engineers working on ScalienDB. Testing basically occured at these clients. Unfortunately, in 2012 Scalien went out of business and ScalienDB never saw production usage after that, nor is it maintained anymore. Below is a short list of issues encountered during production use.

\textbf{Shard migration.} Shard migration, although in the codebase, was never seriously tested. Shard migration is problematic, especially if it occurs during other bulk operations such as catchup, or if mixed with transactions. Some of these edge cases may not be handled correctly in the code. Shard migration is not automatic: there is no code in ScalienDB to automatically move shards from one quorum to another to re-balance, we recommended that our clients do this manually using the API, where this functionality was exposed. The development of this mechanism was delayed because we were afraid of deploying such automatism into production code at that point.

\textbf{Lack of data manipulation language and indices.} Clients complained that examining data in ScalienDB is hard, due to the simplistic key-value data model. Unlike in SQL, where \texttt{SELECT} and \texttt{UPDATE} statements can be used to examine and fix the data, with ScalienDB clients were forced to do so by hand. In one occurence the client had to write a .NET program using the ScalienDB library and deploy it to the production server just to fix some data. Clients also had to manually maintain index structures by hard-coding index data into keys. These were clearly a major shortcoming on ScalienDB, and it was part of our plan to upgrade the data model of ScalienDB.

\textbf{Memory fragmentation.} An interesting problem we encountered on production 64-bit Windows machines was memory fragmentation. The client reported that the database gradually slowed down after restart, and in about 6 hours becomes unusably slow. After it was restarted, it was quick again for a while. We collected debug logs but found no memory or other resource leak. Eventually we had to run a profiler in production, since we were unable to reproduce it on test configurations. Examining the profiler we noticed that the \texttt{free()} calls were taking very long, a sign of memory fragmentation. We worked around this by introduing object caches in problematic parts of the code, mostly in the storage engine. Although this was not a "real bug" it was treated as a show stopper by our clients.

\textbf{Stuck network connections.} We had one case when a TCP connection between two shard servers in a quorum got stuck due to a bug in the networking code. This resulted in weird behaviour when the bug occured. If the third shard server was the primary, everything worked fine, since slaves never communicate with each other. But if one of these became the primary, the quorum would get stuck, as these two were unable to communicate, but TotalPaxos requires all nodes, and the controllers never deactivated either because they didn't perceive a problem, since those TCP connections were unaffected by the bug. Once we found this bug we quickly put in place heartbeats on all TCP connections at the application level, and dropped and rebuilt all TCP connections if the heartbeat didn't go through every second. We also fixed the underlying bug, but this case was a clear indicator of the importance of high level guards to protect against low level bugs.

\textbf{Operational metrics.} Our clients' infrastructure engineers taught us is the importance of operational metrics. Upon their request, we exposed numerous metrics from the internals of the database (such as number of requests served per second), and the infrastructure engineers created dashboards from these. This proved to be an invaluable tool for detecting bugs and performance bugs: if the infrastructure engineers saw something odd in the metrics, they would alert us and if we could not explain the phenomenon as normal operational behaviour, we would start to examine the code and start to reproduce it. Often we eventually found that it was normal behaviour, but very often the metrics did expose bugs. We called this \textit{statistical debugging}, because it did not pinpoint a specific bug as it occured, but indicated the presence of the bug using statistics.

\textbf{Correctness assumption.} Although a lot of effort was put into the replication and the storage engine, and data was never lost in production, there were bugs in replication. This proved to be difficult to fix, because the underlying assumption of Paxos is that the local databases never diverge, so the ScalienDB code had no code paths to fix inconsistencies that might arise from software bugs. This is a shortcoming we realized late in the development process. In retrospect this is a (minor) argument for eventual consistent replication, where such code paths are naturally in the code, and are easier to invoke to fix inconsistencies due to software bugs. Of course the real solution in both cases is large-scale testing of the database to eradicate such bugs in the code.

\textbf{Operational management tools.} In the process of deploying ScalienDB to production, we realized the importance to give infrastructure engineers tools to detect and recover from unexpected errors (including ones caused by software bugs in the database that should not have occured). The most popular part of ScalienDB was the web management console, because it was more complete than that of the competition. This was odd, because it was not part of the main codebase, and never seemed that important to us. But this points to a more general point we missed during development: the importance of the part of the product that is actually seen and used by the clients, such as the client APIs, the command line parameters, logging, metrics, management tools. These are all components not part of the core, which is where most of our efforts went. In retrospect we should have assigned higher importance to these parts of the product, since this is what the client actually sees and experiences.

ScalienDB was a commercial failure, because Scalien was a startup failure. Shortcomings and bugs in ScalienDB were not the root cause of this failure. The root cause was that (i) Scalien never secured venture funding, and (ii) Scalien did not follow lean principles \cite{LeanStartup}, and did not go through quick build-measure-learn cycles to approach product-market fit. It is probably impossible to achieve quick build-measure-learn cycles while writing a monolithic C++ database application. In retrospect, we would pick a platform and architecture that would allow for quick product iterations. For example, use node.js as an application container and write the database in Javascript, and not worry about speed until the product-market fit is validated.

\end{document}